\documentclass[sigconf]{acmart}
\AtBeginDocument{%
  }

\copyrightyear{2025}
\acmYear{2025}
\setcopyright{acmlicensed}
\acmConference[SIGIR '25]{Proceedings of the 48th International ACM SIGIR Conference on Research and Development in Information Retrieval}{July 13--18, 2025}{Padua, Italy}
\acmBooktitle{Proceedings of the 48th International ACM SIGIR Conference on Research and Development in Information Retrieval (SIGIR '25), July 13--18, 2025, Padua, Italy}
\acmDOI{10.1145/3726302.3730042}
\acmISBN{979-8-4007-1592-1/2025/07}

\usepackage{multirow}
\usepackage{booktabs}
\usepackage{color}
\definecolor{lightgray}{RGB}{215,215,215}
\usepackage{colortbl}
\usepackage{xcolor}
\usepackage{subcaption}
\usepackage{footnote}
\usepackage{enumitem}
\usepackage{balance}

\newcommand{\ie}{\emph{i.e., }}
\newcommand{\eg}{\emph{e.g., }}

\newcommand{\wrt}{\emph{w.r.t. }}

\begin{document}

\title{Multi-Grained Patch Training for Efficient LLM-based Recommendation}

\author{Jiayi Liao}
\email{ljy0ustc@mail.ustc.edu.cn}
\orcid{0009-0006-7998-8462}
\affiliation{%
  \institution{University of Science and Technology of China}
  \city{Hefei}
  \country{China}
}

\author{Ruobing Xie}
\authornote{Corresponding authors.}
\email{xrbsnowing@163.com}
\orcid{0000-0003-3170-5647}
\affiliation{%
  \institution{Machine Learning Platform Department, Tencent}
  \city{Beijing}
  \country{China}
}

\author{Sihang Li}
\email{sihang0520@gmail.com}
\orcid{0009-0009-8986-7965}
\affiliation{%
  \institution{University of Science and Technology of China}
  \city{Hefei}
  \country{China}
}

\author{Xiang Wang}
\email{xiangwang1223@gmail.com}
\orcid{0000-0002-6148-6329}
\affiliation{%
  \institution{University of Science and Technology of China}
  \city{Hefei}
  \country{China}
}

\author{Xingwu Sun}
\email{sunxingwu01@gmail.com}
\orcid{0009-0008-3222-0901}
\affiliation{%
  \institution{Machine Learning Platform Department, Tencent}
  \city{Beijing}
  \country{China}
}

\author{Zhanhui Kang}
\email{kegokang@tencent.com}
\orcid{0009-0006-5151-4222}
\affiliation{%
  \institution{Machine Learning Platform Department, Tencent}
  \city{Shenzhen}
  \country{China}
}

\author{Xiangnan He}
\authornotemark[1]
\email{xiangnanhe@gmail.com}
\orcid{0000-0001-8472-7992}
\affiliation{%
  \institution{MoE Key Lab of BIPC, University of Science and Technology of China}
  \city{Hefei}
  \country{China}
}

\settopmatter{authorsperrow=4}
\renewcommand{\shortauthors}{Jiayi Liao et al.}

\begin{abstract}
Large Language Models (LLMs) have emerged as a new paradigm for recommendation by converting interacted item history into language modeling.
However, constrained by the limited context length of LLMs, existing approaches have to truncate item history in the prompt, focusing only on recent interactions and sacrificing the ability to model long-term history. 
To enable LLMs to model long histories, we pursue a concise embedding representation for items and sessions.
In the LLM embedding space, we construct an item's embedding by aggregating its textual token embeddings; similarly, we construct a session's embedding by aggregating its item embeddings.
While efficient, this way poses two challenges since it ignores the temporal significance of user interactions and LLMs do not natively interpret our custom embeddings.
To overcome these, we propose PatchRec, a multi-grained patch training method consisting of two stages: (1) Patch Pre-training, which familiarizes LLMs with aggregated embeddings -- patches, and (2) Patch Fine-tuning, which enables LLMs to capture time-aware significance in interaction history.
Extensive experiments show that PatchRec effectively models longer behavior histories with improved efficiency.
This work facilitates the practical use of LLMs for modeling long behavior histories. 
Codes are available at \url{https://github.com/ljy0ustc/PatchRec}.
\end{abstract}




\maketitle

\section{Introduction}
\begin{figure}[t]
\centering
\includegraphics[width=0.48\textwidth]{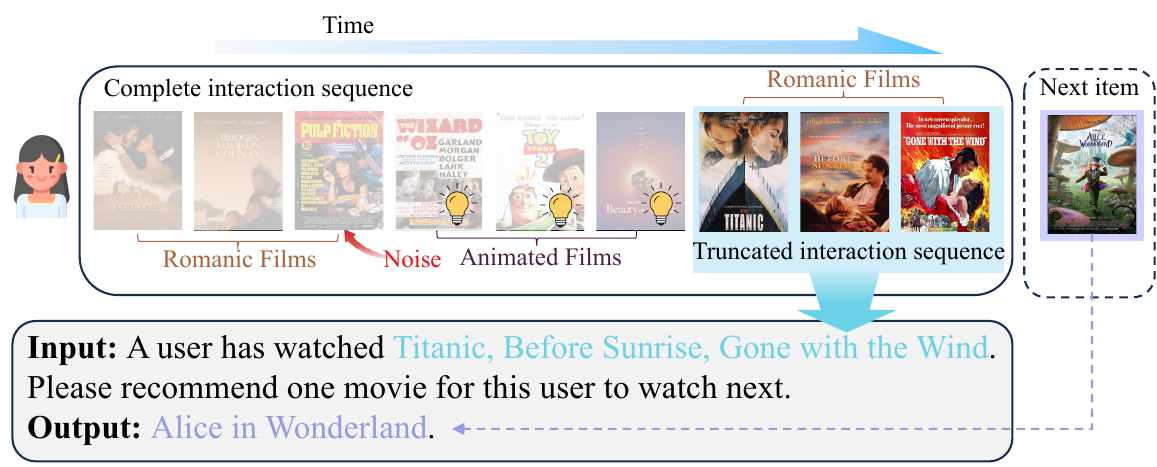}
\caption{Most LLM4SR works consider the latest-$K$ interactions in the complete user interaction sequence as the truncated historical item sequence fed to LLMs.}
\label{fig:teaser}
\end{figure}

Large Language Models (LLMs) \cite{gpt-4, gemini, llama-3, qwen2, deepseek} have demonstrated remarkable success across diverse domains, sparking a growing interest in their application to recommendation. 
A common paradigm for using LLMs as recommender models (LLM4Rec) \cite{p5, llamarec, instructrec, tallrec} involves two key steps: (1) formatting a user's interaction history (\ie a sequence of previously interacted items) into an input prompt suitable for LLMs, and (2) fine-tuning the LLMs to generate the target response (\ie predicting the next item of interest).
For the first step, most leading methods \cite{tallrec, llara, rosepo, recinterpreter} truncate each interaction history, retaining only the most recent $K$ items\footnote{K, with common values being 5, 10, or 20, is typically much smaller than the total length of complete interaction history.}, with each item represented by the textual metadata (\eg movie titles like ``Gone with the Wind'') \cite{tallrec, bigrec, rosepo}, as shown in Figure \ref{fig:teaser}.
This truncation integrates a much shorter interaction sequence into the input prompt, primarily due to LLMs' context window size limitations and the associated computational costs.
While effective for capturing short-term preferences, this approach abandons the interactions prior to the selected items, failing to capture the long-term interests encoded in extended item sequences. 

To efficiently model long item sequences in LLM4Rec, we introduce a hierarchical representation approach, as depicted in Figure \ref{fig:compression}.
In the embedding space, the \textbf{textual tokens} of item titles are firstly compressed into compact \textbf{item patches}.
These items are then organized into sessions, with the corresponding item patches further compressed into \textbf{session patches}, capturing higher-level patterns in user behavior. 
This compressed representation allows for the accommodation of a significantly larger number of items within a fixed context window. 
However, the following two challenges remain:
\begin{itemize}[leftmargin=*]
    \item Temporal significance of interactions:
    Such compressed representations do not explicitly regard the temporal distinctions of relevance \cite{shan}, treating all items in a user's interaction sequence as equally important \cite{llara, bigrec}.
    How can we incorporate mechanisms to capture the varying importance of interactions over time, thereby aligning more closely with evolving user preferences?
    \item Comprehension of compression patterns:
    While compressed item and session patches retain the same dimensionality as original textual tokens, LLMs lack inherent understanding of them.
    How can we design training approaches to enable LLMs to interpret these compressed patterns effectively?
\end{itemize}

To tackle these challenges, we propose \textbf{PatchRec}, a multi-grained patch training framework that unifies: (1) session patches for high-level behavioral patterns, (2) item patches for fine-grained preference modeling, and (3) raw textual tokens for semantic grounding.
This framework enables compact yet expressive representation of user interaction history, while maintaining dynamic adaptability to varying sequence lengths.
As illustrated in Figure~\ref{fig:training}, PatchRec operates in the following two stages:
\begin{itemize}[leftmargin=*]
    \item \textbf{Patch Pre-training}. 
    During each training step, the model simultaneously learns both the original uncompressed and compressed versions of the same data sample, thereby establishing a correspondence between the compressed item patches and their original textual tokens.
    \item \textbf{Patch Fine-tuning}.
    This stage models the temporal variation in item importance using distinct compression granularities. 
    Specifically, items from earlier interactions are compressed to a higher degree, while more recently interacted items are represented by a lower degree of compression, allowing the model to effectively capture longer user histories with varying levels of importance.
\end{itemize}

\begin{figure}[t]
\centering
\includegraphics[width=0.48\textwidth]{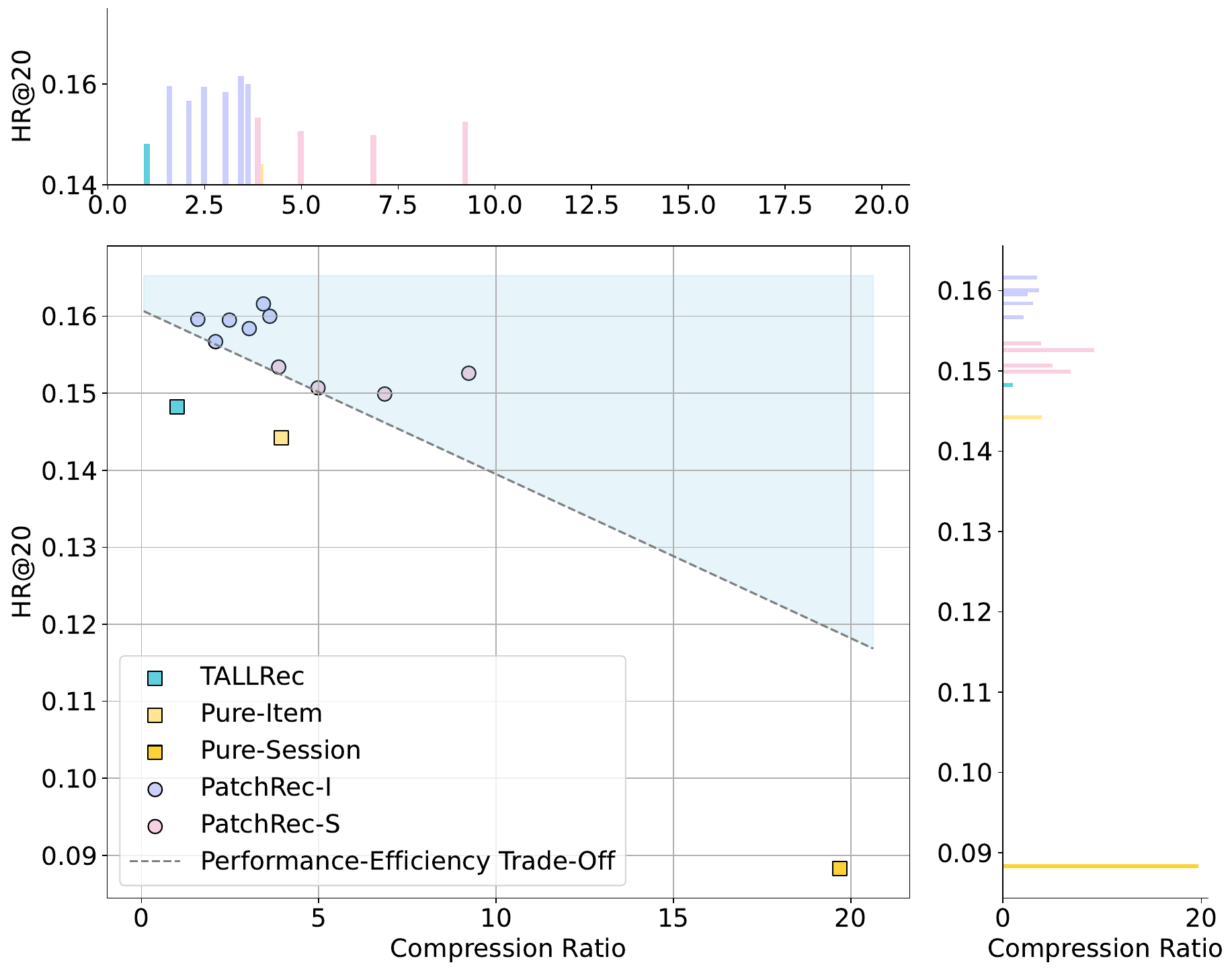}
\vspace{-16pt}
\caption{Performance-efficiency trade-off for TALLRec, Pure-Item/Session, and PatchRec-I/S on MovieLens-1M. PatchRec variants demonstrate improvements in both metrics compared to the baseline. The shaded region highlights the area above the performance-efficiency trade-off curve, representing desirable configurations.}
\vspace{-4pt}
\label{fig:ft2_ml1m}
\end{figure}

Extensive experiments are conducted on datasets and yield the following key findings:
(1) For the same truncated interaction sequence, PatchRec uses only 7.34\% LLM input tokens compared to TALLRec, while simultaneously improving HR@20 by up to 32\% on the Goodreads dataset.
(2) For the same computational cost, PatchRec models 3.44 times more user behaviors than TALLRec, resulting in a substantial performance improvement of up to 13\% on the MovieLens-1M dataset.

Our contributions can be summarized as follows:
\begin{enumerate}[leftmargin=*]
    \item We propose a multi-grained (\ie both item- and session-level)  patch training framework --- PatchRec, modeling users' long-term historical behaviors in LLM4Rec.
    \item Our proposed pre-training and fine-tuning framework enables LLMs to internalize compressed representations as well as time-aware preferences for sequential modeling.
    \item Experimental results demonstrate that our approach not only saves computational resources but also enhances recommendation performance.
\end{enumerate}
\section{Preliminary}
Given an input prompt describing a user's chronological interaction history with items, an LLM-based recommender predicts the next item that best matches this user's preference.
Building upon a foundational work in LLM-based recommendation, TALLRec \cite{tallrec}, we formalize this baseline as supervised finetuning an LLM parameterized by $\Theta$ to maximize the autoregressive likelihood:
\begin{equation}
    P(Y|X;\Theta) = \prod_{t=1}^{T} P(\mathbf{y}_t | \mathbf{x}_1, \mathbf{x}_2, \ldots, \mathbf{x}_{N}, \mathbf{y}_1, \mathbf{y}_2, \ldots, \mathbf{y}_{t-1}; \Theta),
\end{equation}
where $X = x_{1:N}$ is the LLM input sequence of $N$ tokens representing the user interaction history, and $Y = y_{1:T}$ is the output sequence denoting the next item.
The input sequence $X$ encodes a truncated interaction sequence of $K$ items using a structured prompt template. 
Each item is represented through its title, which is subsequently tokenized into several tokens $\mathbf{x}_i \in \mathbb{R}^d$, where $d$ is the hidden size of the LLM.
To adapt TALLRec from click-through rate prediction (answering YES/NO given a target item) to next-item recommendation, we let the LLM generate the next item's title $Y$ token-by-token through autoregressive decoding, employing constrained beam search over a token-level prefix tree constructed from all candidate item titles.
This ensures that generated sequences correspond to valid items.

\begin{figure}[t]
\centering
\includegraphics[width=0.48\textwidth]{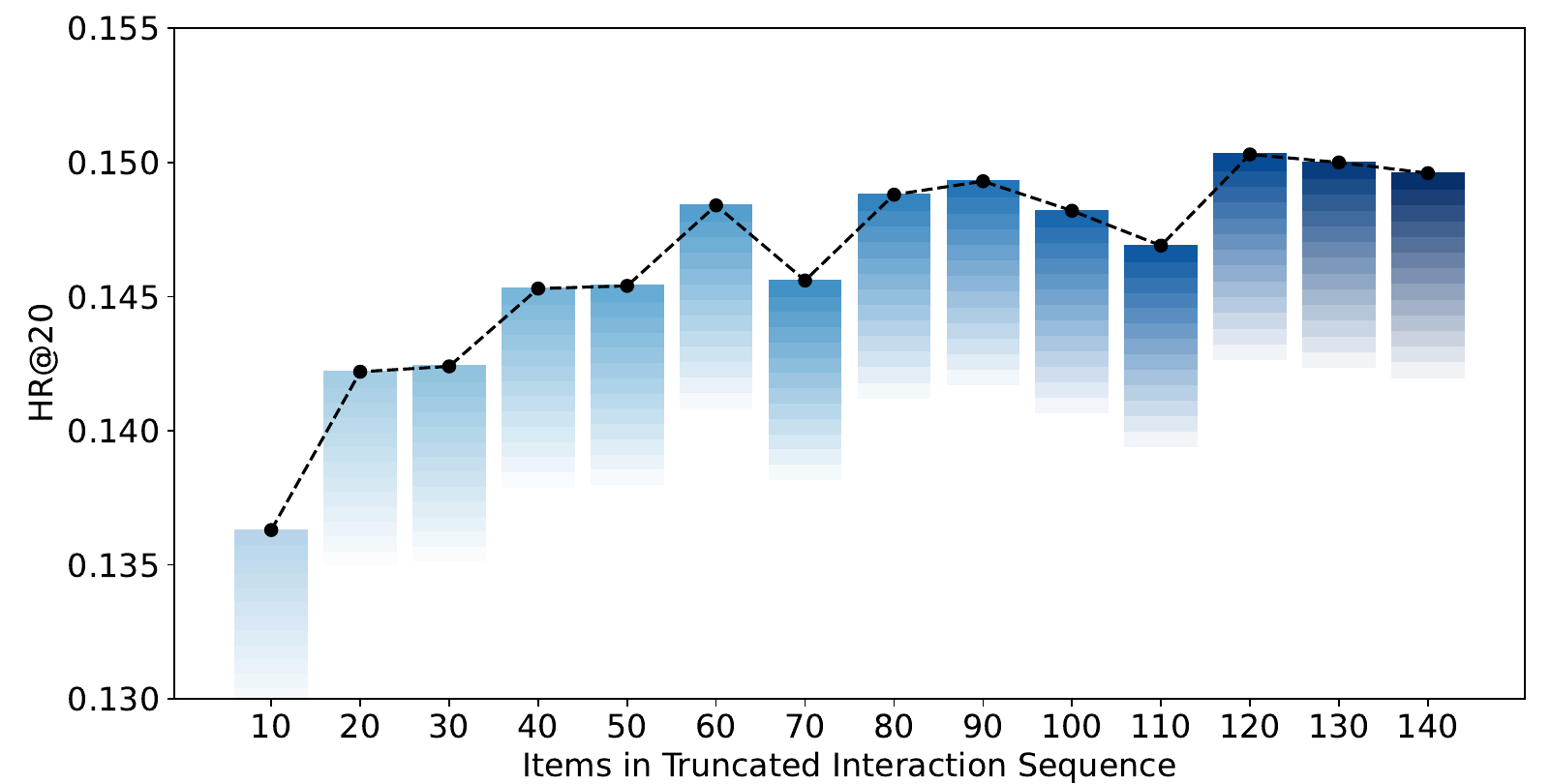}
\caption{Performance of LLM-based recommender on Movielens-1M at different numbers of items in the truncated interaction sequence. The number of tokens in the historical item sequence ranges from 39 to 552, excluding the tokens of task description.}
\vspace{-10pt}
\label{fig:preliminary}
\end{figure}

To investigate the impact of sequence length, we scale up $K$ in preliminary experiments.
Figure \ref{fig:preliminary} reveals a key trend: performance improves with truncated sequence length.
As the number of items in the truncated interaction sequence increases, recommendation accuracy --- HR@20, improves substantially, suggesting that richer user histories enable more precise modeling of preferences.
For instance, earlier interactions with animated films contribute to accurately predicting subsequent items like ``Alice in Wonderland''.

Despite the benefits of modeling long sequences, this baseline approach faces a critical limitation: excessive inference costs due to rapidly growing LLM prompt lengths.
The length of the item sequence significantly influences the input prompt length, which is constrained by the context window length of LLMs \cite{rope} and the associated computational costs.
\section{Method}
In this section, we first present the cornerstone of our approach --- the hierarchical patching strategy --- and then describe the two-stage PatchRec framework.

\subsection{Hierarchical Patching}
\begin{figure}[t]
\centering
\includegraphics[width=0.48\textwidth]{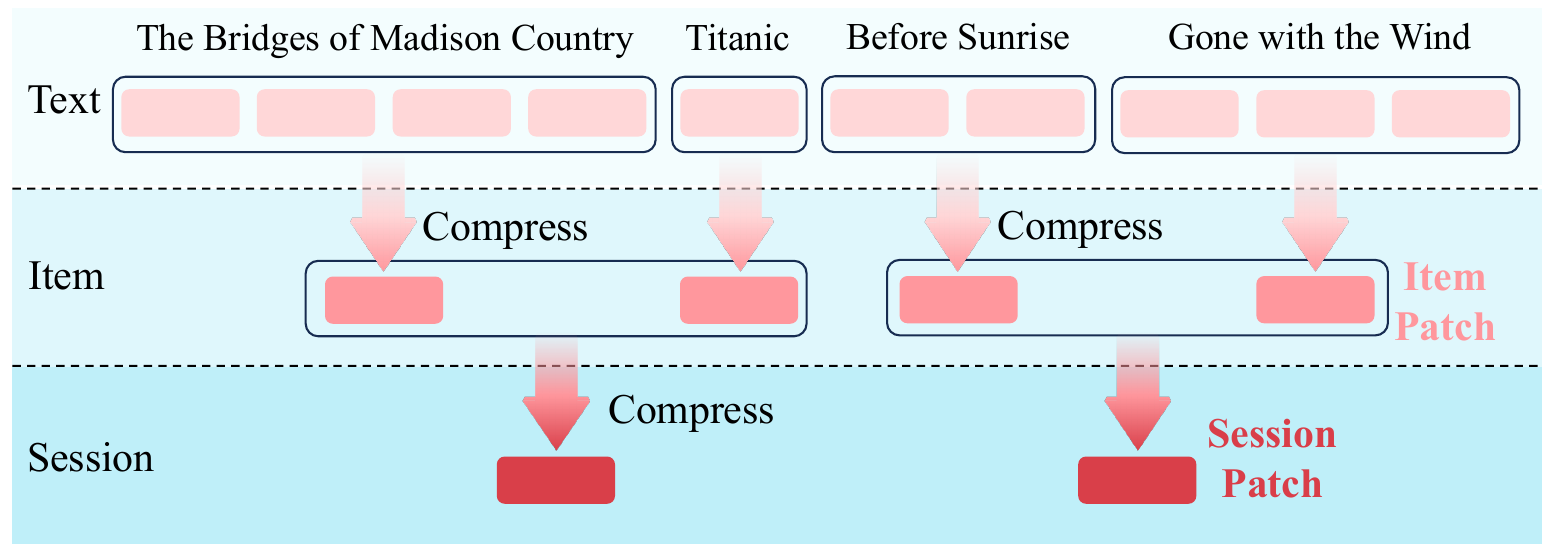}
\vspace{-16pt}
\caption{Hierarchical compression. The textual tokens of an item title are aggregated into a compact item patch. Then, several adjacent item patches are further compressed into a denser session patch.}
\vspace{-4pt}
\label{fig:compression}
\end{figure}

To incorporate more interactions within constraints of LLM prompt length overhead, we hierarchically compress the representation of historical item sequences, as shown in Figure \ref{fig:compression}.

\subsubsection{\textbf{Item Patch Construction.}}
In the context of LLM4Rec, the textual tokens of each item title naturally form a semantically coherent unit. 
To compress these tokens into a single token representation -- termed an item patch -- we adopt a simple yet effective method of average pooling. 
Specifically, we compute the mean of the token embeddings corresponding to the item's title as follows:
\begin{equation}
    \mathbf{z}^i_j = \frac{1}{|\mathcal{T}_j|} \sum_{t \in \mathcal{T}_j} \mathbf{x}_t,
    \label{equ:item-patch}
\end{equation}
where $\mathbf{z}^i_j \in \mathbb{R}^d$ is the resulting item patch for the $j$-th item, $\mathcal{T}_j$ denotes the set of tokens in the item's title, and $\mathbf{x}_t$ is the embedding of token $t$. 
This average pooling approach compresses token-level information into a compact item-level representation.
As shown in Figure \ref{fig:compression}, a movie title such as ``The Bridges of Madison Country'' is encoded into a single item patch. 
This reduces the input length while maintaining the core content of the item description.

\subsubsection{\textbf{Session Patch Construction.}}
Since users often exhibit consistent interests over a period of time, consecutive interactions can be further compressed.
To accomplish this, we partition a user's historical item sequence into sessions, where each session is formed by grouping a fixed number $L$ of consecutive items \footnote{Alternative partitioning methods can also be employed, such as grouping items interacted with during a specific time window.}.
We then obtain a session patch by averaging the embeddings of the item patches within each session:
\begin{equation}
    \mathbf{z}^s_j = \frac{1}{|\mathcal{S}_j|} \sum_{t \in \mathcal{S}_j} \mathbf{z}^i_t,
    \label{equ:session-patch}
\end{equation}
where $\mathbf{z}^s_j \in \mathbb{R}^d$ denotes the session patch for the $j$-th session, $\mathcal{S}_j$ is the set of item patches associated with this session, and $\mathbf{z}^i_t$ represents the embedding of the $t$-th item patch. 
As depicted in Figure \ref{fig:compression}, every two adjacent item patches are further compressed into a denser session patch. 
This hierarchical compression strategy enables efficient handling of longer sequences while preserving the underlying user interest patterns.

\subsubsection{\textbf{Challenges of Patch Training}}
Based on our proposed item and session patch compression methods, we first examine two intuitive adaptations: substituting all text tokens in a sequence with item or session patches only, named Pure-Item and Pure-Session.
While effectively reducing LLM input tokens, they ignore the temporal significance of user interactions and the LLM's inherent incomprehension of compression patterns, suggesting clear opportunities for improvement.
As shown in Figure~\ref{fig:ft2_ml1m}, these approaches achieve suboptimal performance-efficiency trade-offs, remaining below the Pareto frontier in the upper-right optimization space.

\subsection{PatchRec}
\begin{figure*}[t]
\centering
\includegraphics[width=0.95\textwidth]{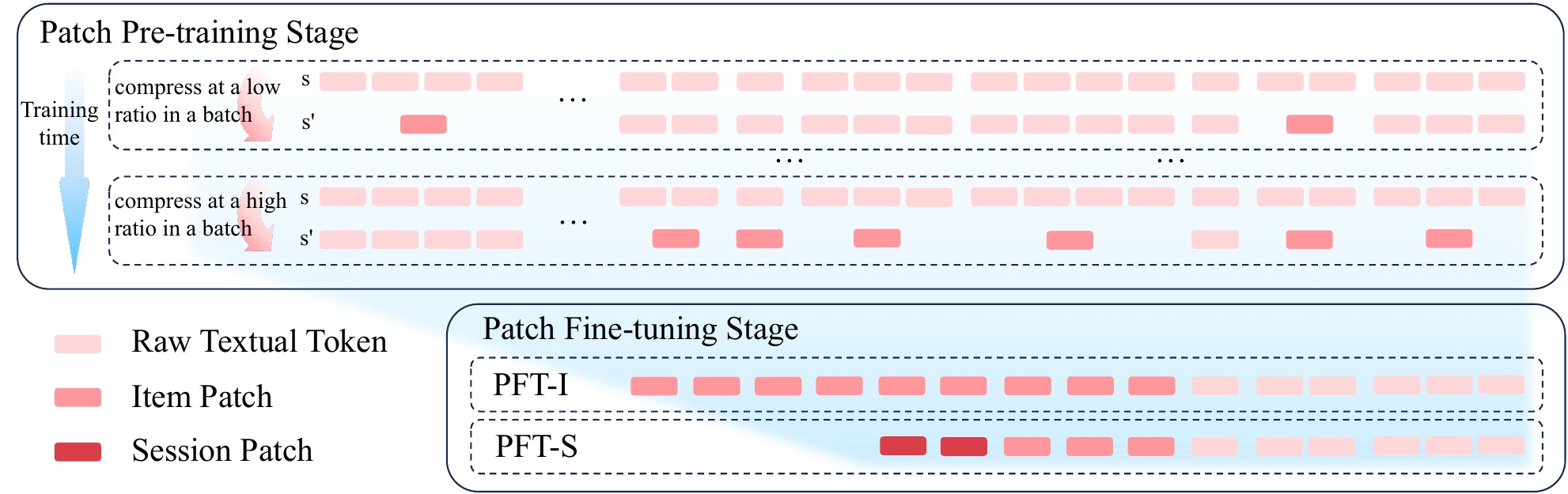}
\caption{Two-stage training framework of PatchRec. In the patch pre-training stage, we augment each uncompressed sequence $s$ with a compressed version $s'$, where each item is independently and randomly compressed from raw textual tokens into an item patch, in order to build connections between item patches and textual item titles. During the training process, the degree of compression gradually increases from 0 to 1. In the patch finetuning stage, we fine-tune the LLMs with time-aware compressed sequences, allowing the LLM to become familiar with a mixed space of various compression granularities in downstream usage. For each sequence, interactions that occur earlier are compressed to a greater extent, while interactions that occur later are compressed to a lesser extent. PFT-S can achieve higher compression ratio with denser session patches.}
\label{fig:training}
\end{figure*}

To accommodate both challenges of time-aware sequence modeling and compression pattern understanding, we propose \textbf{PatchRec}, a simple yet effective framework featuring two-stage training, as illustrated in Figure \ref{fig:training}.
To model the varying relevance of earlier versus later interacted items to a user's current interests, we employ a multi-grained patching strategy. Specifically, earlier interactions are compressed at higher levels (\eg as item and even session-level patches) for capturing long-term preferences, while most recent interactions retain finer granularity (\eg textual tokens) for preserving semantic grounding.
Since LLMs do not inherently understand the compression pattern from textual tokens to item patches, in the first stage, Patch Pre-training, we train the LLM to learn it by comparing item patches with their corresponding original textual tokens. 
In the second stage, Patch Fine-tuning, we further adapt the LLM to the final multi-grained compressed format.

\subsubsection{\textbf{Patch Pre-training}}
For each interaction sequence in a training batch, as shown in Figure \ref{fig:training}, where items are represented by their textual titles, we augment the interaction sequence with a compressed version of it.
In this compressed interaction sequence, the textual tokens of each item are independently and randomly compressed into an item patch with probability $p$. 
The training objective is formulated as:
\begin{equation}
    P(Y|X;\Theta) = P(Y| \mathbf{x}_1, \mathbf{x}_2, \ldots, \mathbf{x}_{k-1},\mathbf{z}^i_j,\mathbf{x}_{k+n+1}, \ldots,\mathbf{x}_N; \Theta),
\end{equation}
where the $j$-th item' title tokens $\mathbf{x}_{[k:k+n]}$ are compressed into an item patch $\mathbf{z}^i_j$ as defined in Equation \ref{equ:item-patch}.  
The probability $p$ gradually increases from 0 to 1 during training, following the schedule:
\begin{equation}
    p = \frac{\tau}{T},
\end{equation}
where $\tau$ denotes the current training step, and $T$ represents the total number of training steps. 
By employing $p$ as a scheduler, the item representations in the compressed version progressively transition from textual tokens to item patches over the course of training.

Within each batch, interaction sequences in both raw and compressed formats are presented simultaneously. 
This approach establishes a clear correspondence for each interaction, effectively linking the textual tokens to their respective item patches.

\subsubsection{\textbf{Patch Fine-tuning}}
To incorporate the temporal decay of user interests in a sequence, we propose two fine-tuning strategies with different compression granularities: \underline{P}atch \underline{F}ine-\underline{t}uning on \underline{I}tems (PFT-I) and \underline{P}atch \underline{F}ine-\underline{t}uning on \underline{S}essions (PFT-S).
As shown in Figue \ref{fig:training}, they are designed to accommodate varying computational resource constraints.

\textbf{PFT-I} retains the most recent $M$ items  \footnote{$M$ is a predefined hyperparameter, which controls the overall compression ratio.} from a truncated interaction sequence of $K$ items, represented by their textual tokens.
The earlier $(K - M)$  items are compressed into item patches. 
The training objective is formulated as:
\begin{equation}
    P(Y|X;\Theta) = P(Y| \mathbf{z}^i_1, \mathbf{z}^i_2, \ldots, \mathbf{z}^i_{K-M},\mathbf{x}_j, \ldots,\mathbf{x}_N; \Theta),
    \label{equ:PFT-I}
\end{equation}
where $\mathbf{x}_j$ denotes the start textual token of the $(K - M+1)$-th item and $N$ is the total number of tokens across all $K$ truncated items.
PFT-I achieves a moderate compression ratio.

\textbf{PFT-S} divides the truncated historical item sequence into groups of items based on interaction time, with each group containing at most $L$ adjacent items. 
The interactions in the latest group are retained as original textual tokens, while those in the second-latest group are compressed into item patches. 
All earlier groups are further compressed into the densest session patches.
The training objective is formulated as:
\begin{equation}
    P(Y|X;\Theta) = P(Y| \mathbf{z}^s_1, \mathbf{z}^s_2, \ldots, \mathbf{z}^i_{1},\mathbf{z}^i_2, \ldots,\mathbf{x}_1,\mathbf{x}_2, \ldots,; \Theta).
\end{equation}
PFT-S achieves a higher compression ratio.

\begin{table*}[h]
\centering
\caption{Statistics of datasets. \textbf{Int.}: Interactions. \textbf{SL.}: Items per user sequence. \textbf{TT.}: Title tokens per item.}
\vspace{-4pt}
\label{tab:data_statis}
\begin{tabular}{lrrrrrr}
\toprule
Dataset& \# User & \# Item & \# Int. & \# Max SL. & \# Avg. SL. & \# Avg. TT. \\
\midrule
MovieLens-1M & 6,037 & 3,883 & 575,281 & 1,435 & 95.28 &  3.94 \\
Goodreads & 2,090 & 8,793 & 162,539 & 430 & 77.77 & 9.82 \\
MovieLens-100K & 682 & 1,682 & 51,824 & 378 & 75.99 & 3.86 \\
\bottomrule
\end{tabular}
\end{table*}
\begin{table*}[t]
\caption{Performance comparison with 100 items in the truncated user sequence.
CR is short for compression ratio.
Bold and underlined indicate the best and the second-best results, respectively. PatchRec improves recommendation accuracy with less computational demands. *(p-value $\ll$ 0.05)}
\setlength{\tabcolsep}{1.75mm}{
\resizebox{\textwidth}{!}{
\begin{tabular}{l|ccccc|ccccc|ccccc}
\toprule
 & \multicolumn{5}{c|}{\textbf{MovieLens-1M*}} & \multicolumn{5}{c|}{\textbf{Goodreads*}} & \multicolumn{5}{c}{\textbf{MovieLens-100K}}\\              
 \textbf{Model} & \textbf{HR@10} & \textbf{N@10} & \textbf{HR@20} & \textbf{N@20} & \cellcolor{cyan!6}\textbf{CR} & \textbf{HR@10} & \textbf{N@10} & \textbf{HR@20} & \textbf{N@20} & \cellcolor{cyan!6}\textbf{CR} & \textbf{HR@10} & \textbf{N@10} & \textbf{HR@20} & \textbf{N@20} & \cellcolor{cyan!6}\textbf{CR} \\ \hline\hline
\textbf{GRU4Rec} & 0.0623 & 0.0298 & 0.1106 & 0.0419 & \cellcolor{cyan!6}- & 0.0401 & 0.0219 & 0.0592 & 0.0267 & \cellcolor{cyan!6}- & 0.0661 & 0.0307 & 0.1211 & 0.0445 & \cellcolor{cyan!6}- \\
\textbf{Caser} & 0.0384 & 0.0180 & 0.0712 & 0.0262 & \cellcolor{cyan!6}- & 0.0116 & 0.0057 & 0.0220 & 0.0083 & \cellcolor{cyan!6}- & 0.0501 & 0.0234 & 0.0910 & 0.0336 & \cellcolor{cyan!6}- \\
\textbf{SASRec} & 0.0594 & 0.0286 & 0.1035 & 0.0397 & \cellcolor{cyan!6}- & 0.0404 & 0.0229 & 0.0569 & 0.0271 & \cellcolor{cyan!6}- & 0.0605 & 0.0290 & 0.1069 & 0.0406 & \cellcolor{cyan!6}- \\
\midrule
\textbf{LinRec} & 0.0622 & 0.0296 & 0.1063 & 0.0407 & \cellcolor{cyan!6}- & 0.0251 & 0.0125 & 0.0398 & 0.0161 & \cellcolor{cyan!6}- & 0.0619 & 0.0283 & 0.1048 & 0.0393 & \cellcolor{cyan!6}- \\
\textbf{Mamba4Rec} & 0.0648 & 0.0303 & 0.1116 & 0.0421 & \cellcolor{cyan!6}- & 0.0281 & 0.0146 & 0.0450 & 0.0188 & \cellcolor{cyan!6}- & 0.0667 & 0.0309 & 0.1086 & 0.0415 & \cellcolor{cyan!6}- \\
\midrule
\textbf{MoRec} & 0.0341 & 0.0164 & 0.0616 & 0.0233 & \cellcolor{cyan!6}- & 0.0122 & 0.0061 & 0.0236 & 0.0089 & \cellcolor{cyan!6}- & 0.0383 & 0.0184 & 0.0762 & 0.0279 & \cellcolor{cyan!6}- \\
\textbf{TALLRec} & 0.0933 & 0.0396 & 0.1482 & 0.0491 & \cellcolor{cyan!6}1.00 & 0.0563 & 0.0236 & 0.0770 & 0.0249 & \cellcolor{cyan!6}1.00 & \underline{0.0986} & \underline{0.0443} & \underline{0.1680} & \textbf{0.0611} & \cellcolor{cyan!6}1.00 \\
\midrule
\cellcolor{gray!16}\textbf{PatchRec-I} & \cellcolor{gray!16}\textbf{0.1058} & \cellcolor{gray!16}\textbf{0.0455} & \cellcolor{gray!16}\textbf{0.1616} & \cellcolor{gray!16}\textbf{0.0525} & 
\cellcolor{cyan!6}\underline{3.44} & \cellcolor{gray!16}\underline{0.0733} & \cellcolor{gray!16}\underline{0.0289} &\cellcolor{gray!16}\underline{0.0976} &\cellcolor{gray!16}\underline{0.0293}
&\cellcolor{cyan!6}\underline{6.81}
& \cellcolor{gray!16}0.0967 & 
\cellcolor{gray!16}0.0416 &
\cellcolor{gray!16}\textbf{0.1738} &\cellcolor{gray!16}\underline{0.0574}
&\cellcolor{cyan!6}\underline{3.38} \\
\cellcolor{gray!16}\textbf{PatchRec-S} & \cellcolor{gray!16}\underline{0.0967} & \cellcolor{gray!16}\underline{0.0408} & \cellcolor{gray!16}\underline{0.1526} & \cellcolor{gray!16}\underline{0.0496} & 
\cellcolor{cyan!6}\textbf{9.23} &
\cellcolor{gray!16}\textbf{0.0748} & \cellcolor{gray!16}\textbf{0.0295} &\cellcolor{gray!16}\textbf{0.1013} &\cellcolor{gray!16}\textbf{0.0301} &
\cellcolor{cyan!6}\textbf{13.62} &
\cellcolor{gray!16}\textbf{0.1016} & \cellcolor{gray!16}\textbf{0.0450} &\cellcolor{gray!16}\underline{0.1680} &\cellcolor{gray!16}0.0554 &
\cellcolor{cyan!6}\textbf{3.85} \\ \bottomrule
\end{tabular}
}}
\label{tab:performance_comparison}
\end{table*}

Fine-tuning the LLM on these multi-grained compressed interaction sequences enables it to effectively capture time-aware user interest patterns, striking a balance between representation effectiveness and computational efficiency.
\section{Experiments}
In this section, we present a comprehensive comparison of PatchRec against various baseline methods across real-world datasets, demonstrating the effectiveness and efficiency of PatchRec under different levels and granularities of compression. 
Additionally, we analyze the mechanisms through which our approach achieves its superior performance and its performance in user groups with more long-term behaviors.

\subsection{Experimental Settings}
\textbf{Datasets.}
We conduct extensive experiments on three real-world benchmark datasets: MovieLens-1M, Goodreads, and MovieLens-100K.
The MovieLens datasets are widely used benchmarks, sourced from a movie recommendation platform \footnote{https://movielens.org/}, while the Goodreads dataset is derived from the largest online community for readers and book recommendations \footnote{https://www.goodreads.com/}.

To ensure data quality, we apply filtering based on rating thresholds of 3 and 5 for MovieLens and Goodreads, respectively.
For MovieLens-100K, we retain only users with more than 20 interactions. 
For Goodreads, we exclude users who have interacted with fewer than 50 items and items that have received fewer than 10 interactions from distinct users.
To prevent information leakage and balance computational overhead during training and evaluation, we partition the interactions in each dataset into training, validation, and test splits, following a temporal split ratio of 48:1:1 based on timestamps.
The dataset statistics are summarized in Table \ref{tab:data_statis}.

\begin{figure*}
    \centering
    \begin{subfigure}[b]{0.33\textwidth}
        \includegraphics[width=\textwidth]{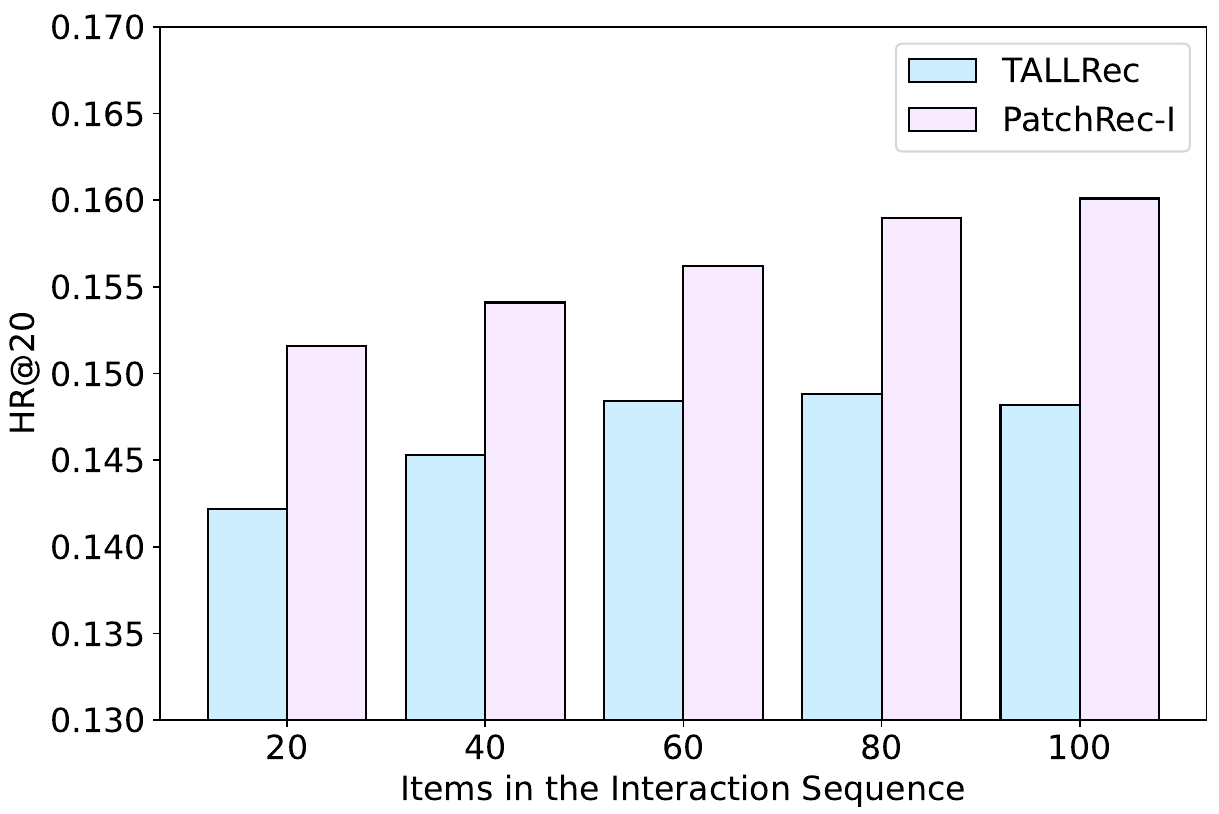}
        \caption{MovieLens-1M}
        \label{fig:ft1-history-ml1m}
    \end{subfigure}
    \begin{subfigure}[b]{0.33\textwidth}
        \includegraphics[width=\textwidth]{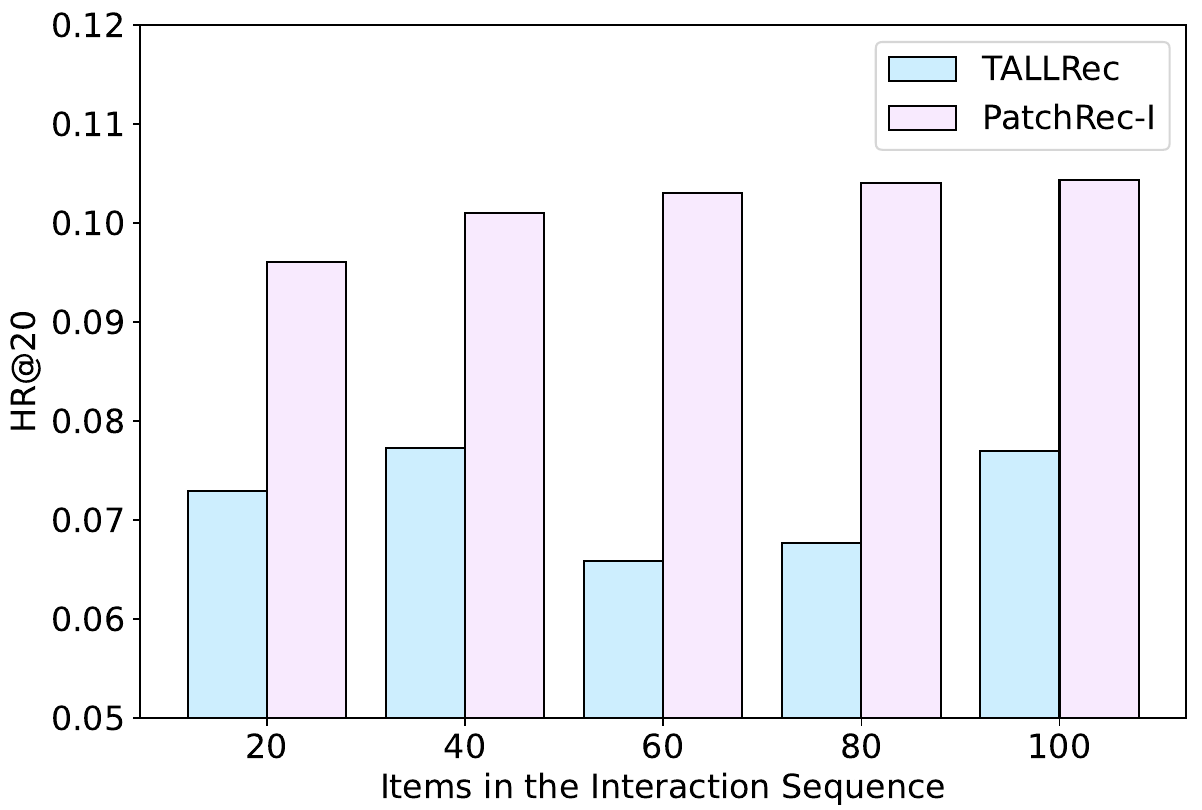}
        \caption{Goodreads}
        \label{fig:ft1-history-goodreads}
    \end{subfigure}
    \begin{subfigure}[b]{0.33\textwidth}
        \includegraphics[width=\textwidth]{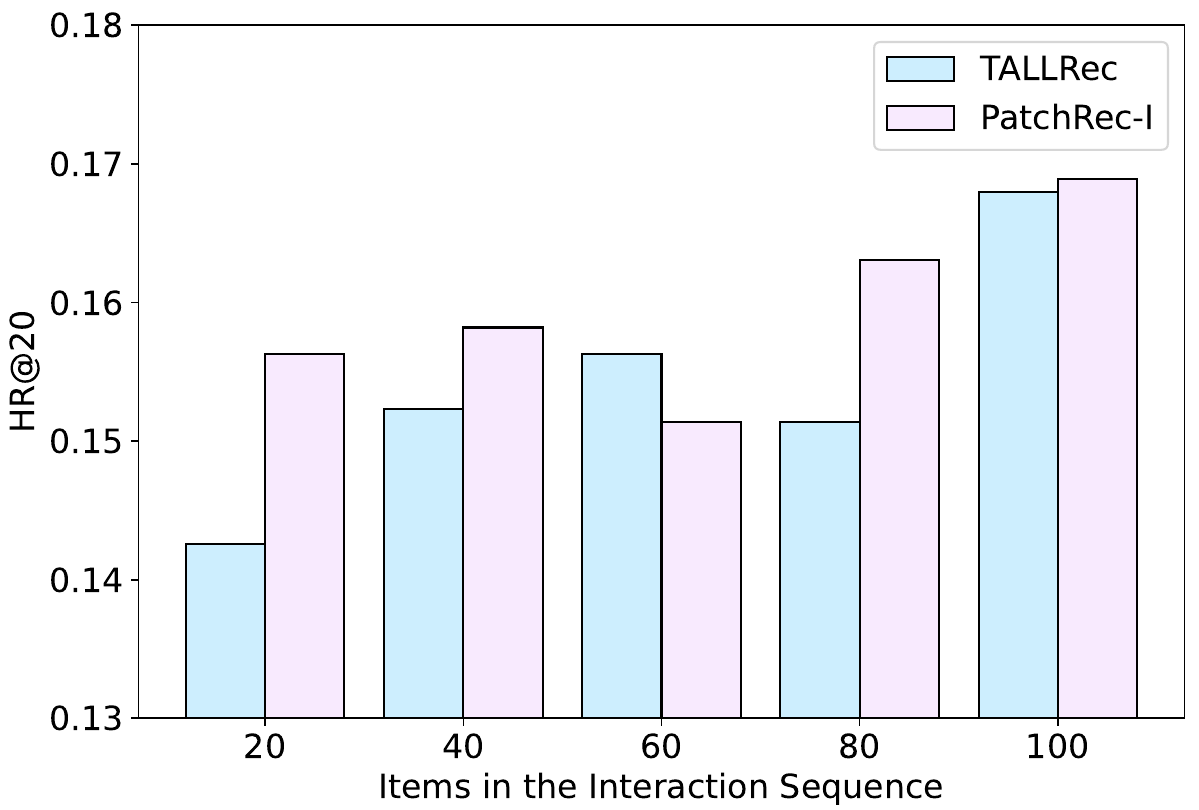}
        \caption{MovieLens-100K}
        \label{fig:ft1-history-ml100k}
    \end{subfigure}
    \vspace{-16pt}
    \caption{Performance comparison between PatchRec-I with TALLRec with the same item numbers in interaction sequence. The compression ratios of PatchRec-I are 2.27, 3.06, and 2.25 for MovieLens-1M, Goodreads, and MoviLens-100K, respectively. Impressively, with improved efficiency, PatchRec-I does not compromise recommendation accuracy.}
    \label{fig:ft1-history}
\end{figure*}

\begin{figure*}
    \centering
    \begin{subfigure}[b]{0.33\textwidth}
        \includegraphics[width=\textwidth]{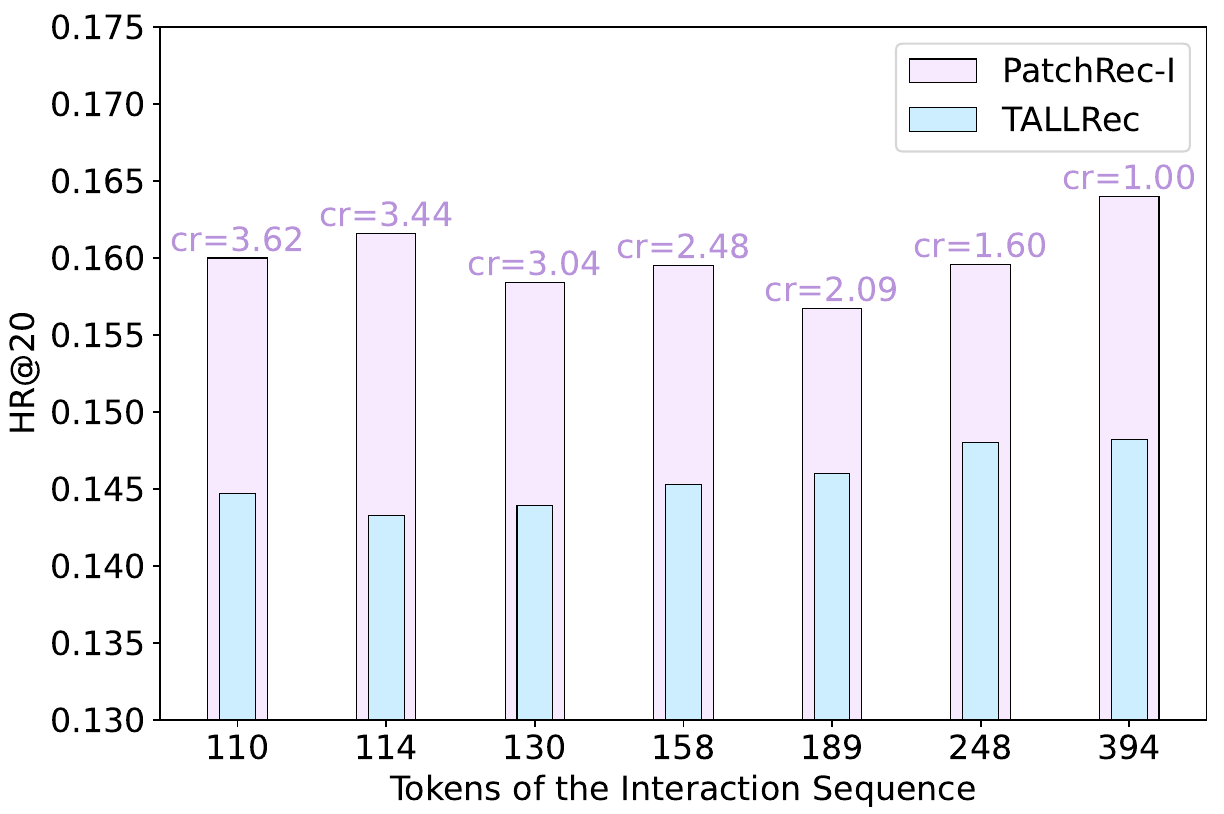}
        \caption{MovieLens-1M}
        \label{fig:ft1-token-ml1m}
    \end{subfigure}
    \begin{subfigure}[b]{0.33\textwidth}
        \includegraphics[width=\textwidth]{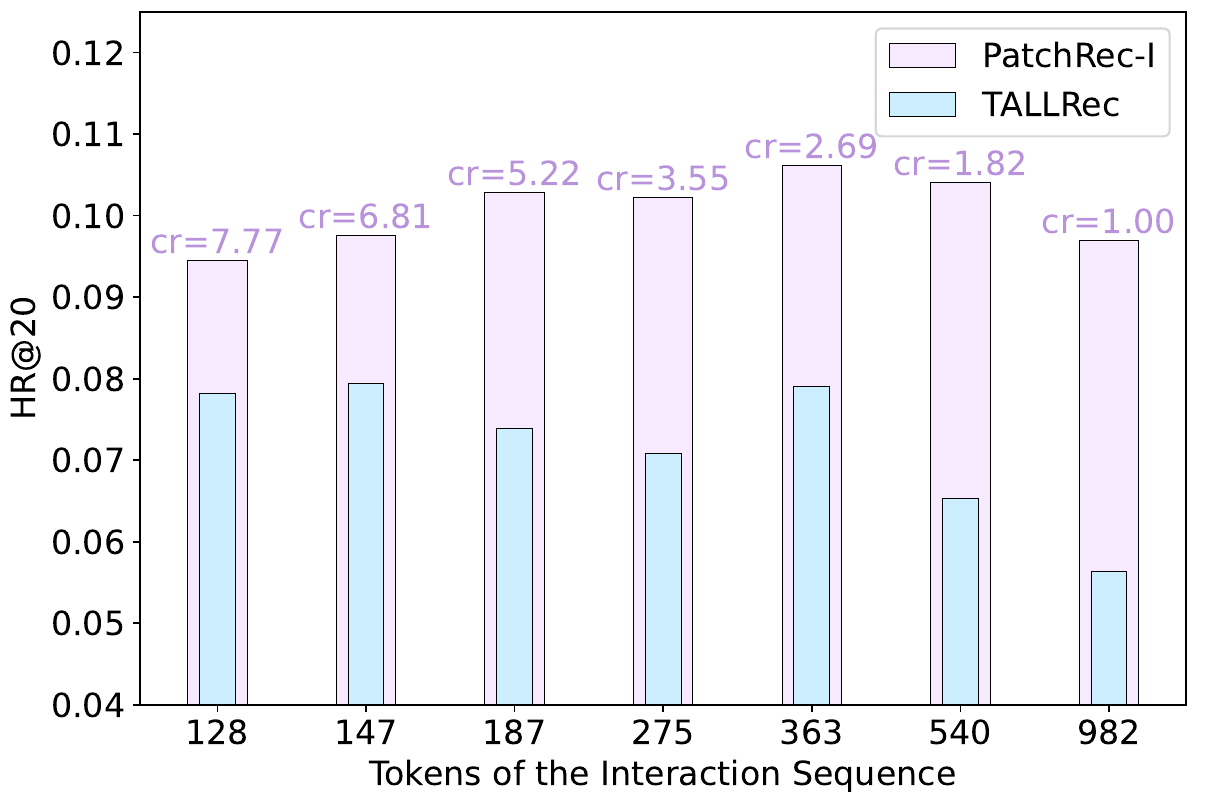}
        \caption{Goodreads}
        \label{fig:ft1-token-goodreads}
    \end{subfigure}
    \begin{subfigure}[b]{0.33\textwidth}
        \includegraphics[width=\textwidth]{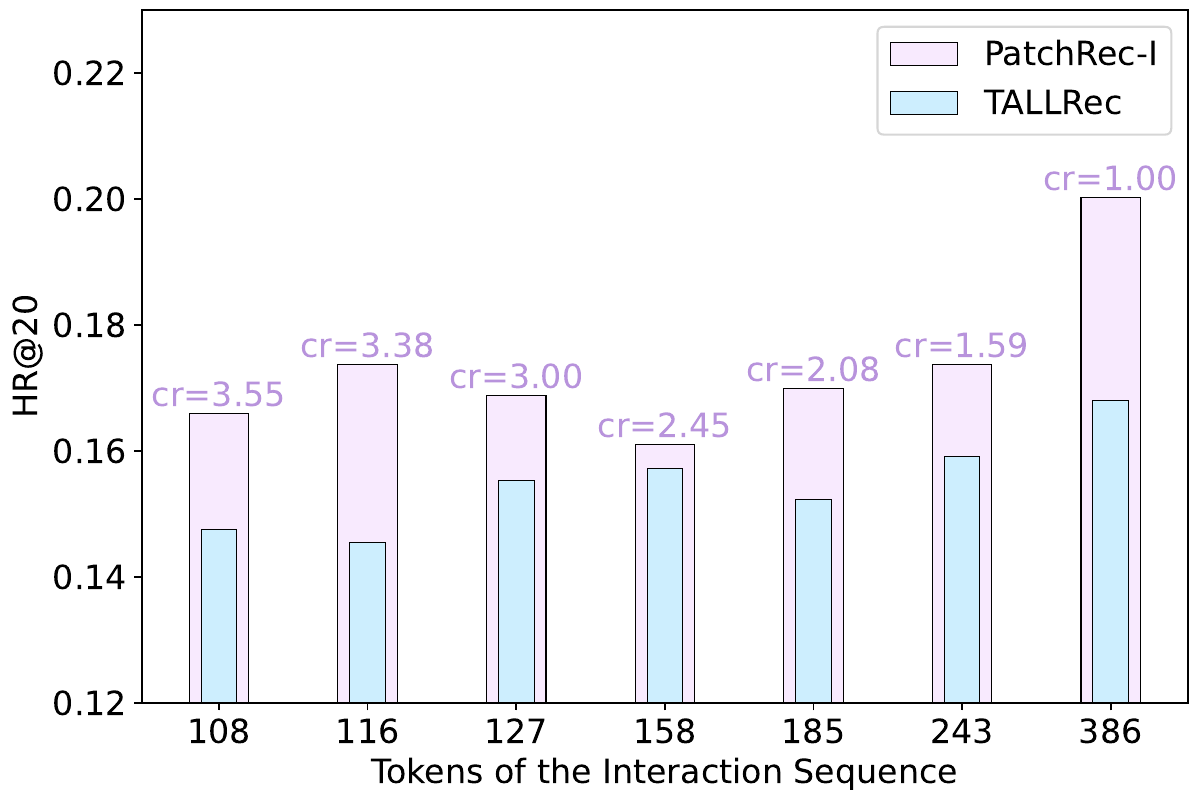}
        \caption{MovieLens-100K}
        \label{fig:ft1-token-ml100k}
    \end{subfigure}
    \caption{Performance comparison between PatchRec-I with TALLRec with comparable numbers of tokens in the interaction sequence. 
    The item number in the interaction sequence is 100 for PatchRec-I, with various compression ratios (cr). 
    For PatchRec-I, the latest-$M$ ($M$=3, 5, 10, 20, 30, 50, 100) items are represented with textual titles, and earlier items are compressed into item patches.
    PatchRec-I delivers superior recommendation performance compared to TALLRec at the same computational cost.}
    \label{fig:ft1-token}
\end{figure*}

\textbf{Implementations.}
We adopt Llama-3.2-1B-Instruct \cite{llama-3} as the backbone LLM, fine-tuning all its parameters. 
Each training stage consists of one epoch.
To balance computational efficiency and resource allocation, we use a training batch size of 64 for MovieLens datasets and 32 for Goodreads, and a test batch size of 16 for all three datasets.
The training employs a cosine learning rate scheduler with a warmup ratio of 0.05, and the learning rates are set to 8e-6,5e-5,1e-5 for MovieLens-1M, Goodreads, and MovieLens-100K, respectively.
For each interaction, the latest $K$ historical interactions are selected as the truncated user sequence. 
All experiments access item titles that appear within their respective datasets to ensure fair evaluation.

\textbf{Evaluation.}
To evaluate recommendation performance, we use two key metrics: HitRatio@10/20 and NDCG@10/20.
Additionally, to assess efficiency, we introduce the \textbf{Compression Ratio}, defined as the ratio of the number of tokens in the historical item sequence before compression to the number after compression. 
A higher compression ratio indicates greater efficiency.

\subsection{Main Performance Comparison}
\subsubsection{\textbf{Baselines}}
We denote the models trained using patch pre-training and PFT-I as \textbf{PatchRec-I}, and those trained with patch pre-training and PFT-S as \textbf{PatchRec-S}.
They are compared against traditional recommender models, efficient long-term sequential recommender models, and LLM-based recommenders.
\begin{itemize}[leftmargin=*]
    \item \textbf{Traditional Recommender Models.} We select three sequential recommender models as representatives of different mechanisms for capturing user behavior patterns: GRU4Rec\cite{gru4rec} (RNN-based), Caser\cite{caser} (CNN-based) and SASRec\cite{sasrec} (attention-based) \footnote{These models are implemented following \cite{dros}.}.
    \item \textbf{Efficient Long-term Recommender Models.} LinRec \cite{linrec} and Mamba4Rec \cite{mamba4rec} are chosen as representatives of efficient long-term sequential recommendation methods. 
    LinRec leverages linear attention mechanisms, while Mamba4Rec employs Mamba blocks \cite{mamba} for efficient modeling of long-term user behaviors. 
    \item \textbf{LLM-based Recommenders.} We include two representative methods that incorporate LLMs in sequential recommendation:  (1) MoRec \footnote{We adopt BERT as the encoder for items' textual metadata and SASRec as the recommender backbone.} \cite{morec} enhances traditional recommenders by incorporating textual features encoded by LLMs.
    (2) TALLRec \cite{tallrec} applies SFT on truncated interaction sequences without compression. \footnote{We adapt TALLRec to the all-ranking task with constrained beam search in item title space, consistent with PatchRec}
\end{itemize}

\begin{figure*}
    \centering
    \begin{subfigure}[b]{0.45\textwidth}
        \includegraphics[width=\textwidth]{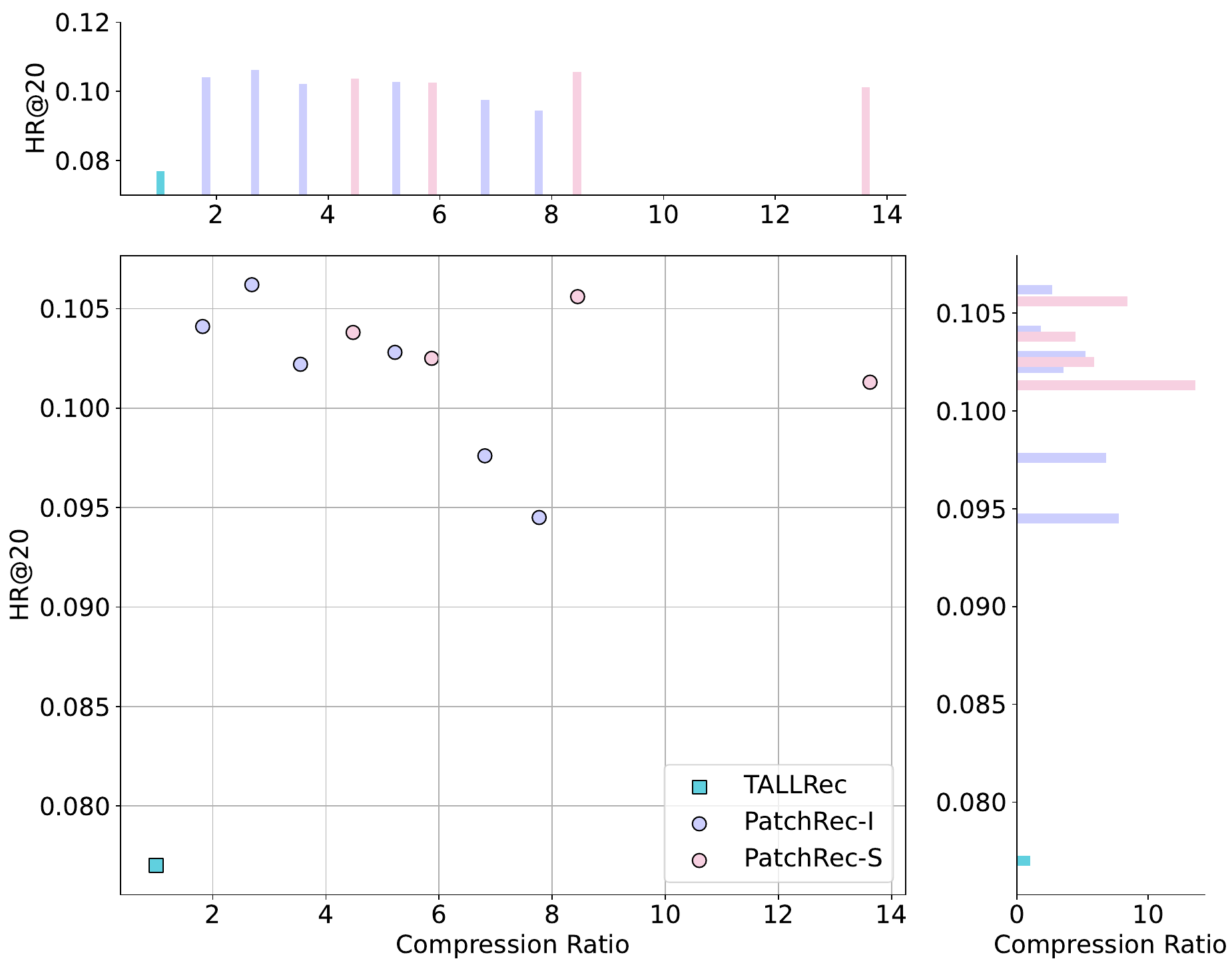}
        \caption{Goodreads}
        \label{fig:ft2-goodreads}
    \end{subfigure}
    \begin{subfigure}[b]{0.45\textwidth}
        \includegraphics[width=\textwidth]{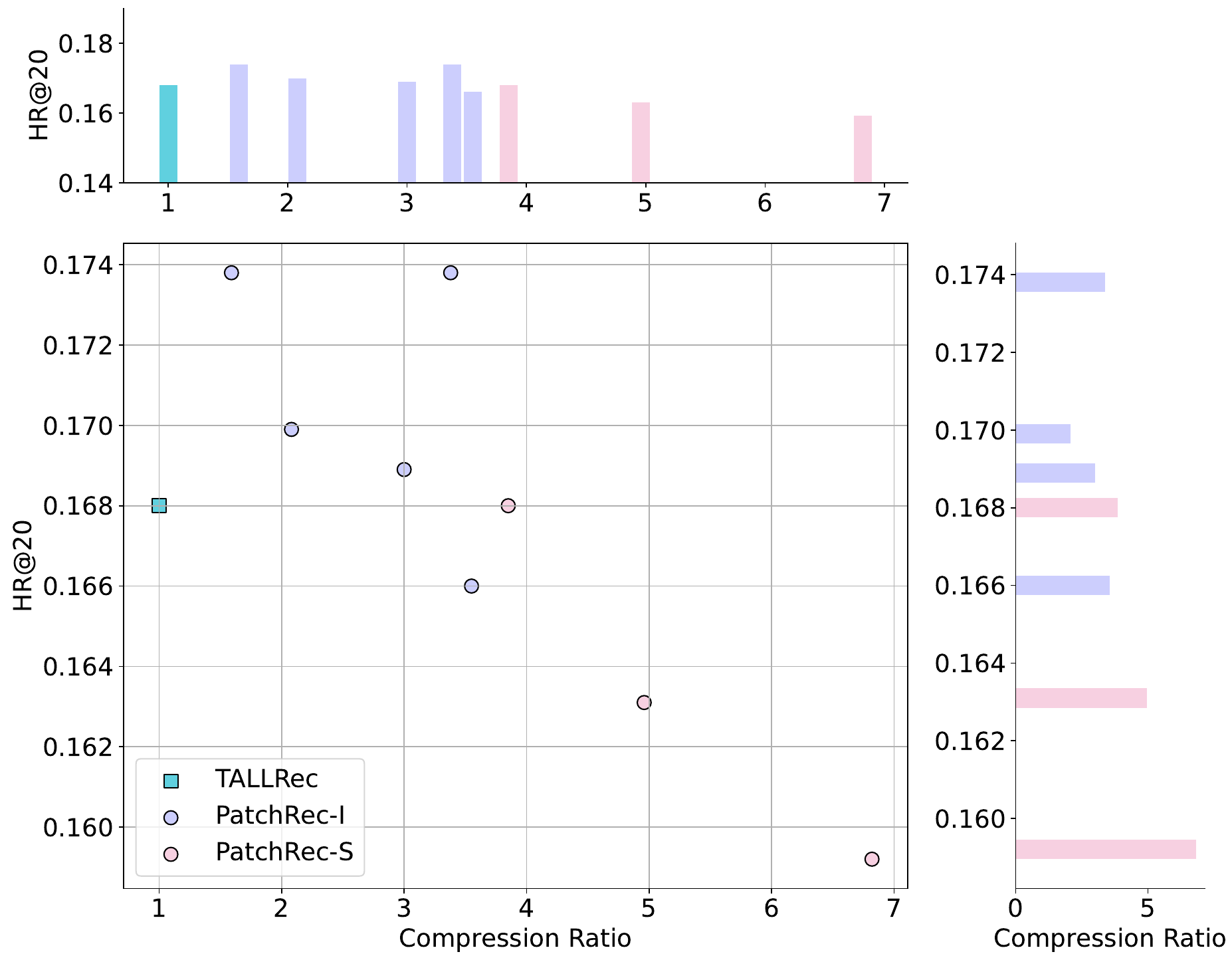}
        \caption{MovieLens-100K}
        \label{fig:ft2-ml100k}
    \end{subfigure}
    \caption{The performance-efficiency trade-off of TALLRec, PatchRec-I, and PatchRec-S on Goodreads and MovieLens-100K dataset. In the scatter plot, points that are positioned further towards the upper right corner indicate better performance and higher efficiency.
    Both PatchRec-I and PatchRec-S provide a better performance-efficiency trade-off compared to TALLRec.}
    \label{fig:ft2}
\end{figure*}

\subsubsection{\textbf{Results}}
The truncation lengths of all the methods compared in Table \ref{tab:performance_comparison} are 100 (\ie modeling the latest-100 items for each sequence).
For PatchRec-I, we keep the latest-$M$ ($M$=5) interactions as items' textual titles and compress the early interactions into item patches.
For PatchRec-S, we use a group size $L$ of 5 for MovieLens-1M and Goodreads, and 20 for MovieLens-100K; so the latest-$L$ interactions are retained as textual tokens, the $L$ interactions in the second latest group are compressed into item patches, and the earlier interactions are compressed into session patches.

The results in Table \ref{tab:performance_comparison} demonstrate that both PatchRec-I and PatchRec-S outperform all traditional recommender, efficient long-term sequential recommender, and LLM-based recommender baselines across all four metrics on MovieLens-1M and GoodReads.
And on MovieLens-100K, with comparable recommendation accuracy, PatchRec achieves a much short input sequence.

Notably, PatchRec-I achieves HR@20 improvements of 9.04\%, 26.75\%, and 3.45\% over TALLRec, with compression ratios of 3.44, 6.81, and 3.38 on MovieLens-1M, Goodreads, and MovieLens-100K, respectively. 
Similarly, PatchRec-S achieves HR@20 improvements of 2.97\% and 31.56\% on MovieLens-1M and Goodreads, respectively, with higher compression ratios of 9.23 and 13.62. 
On MovieLens-100K, PatchRec-S achieves comparable performance to TALLRec with a compression ratio of 3.85.

These results demonstrate that both PatchRec-I and PatchRec-S significantly reduce the number of LLM input tokens without compromising recommendation performance.
Overall, the experimental findings highlight the effectiveness of our approach in simultaneously improving recommendation accuracy and reducing computational overhead.

\subsection{Analysis of PatchRec-I}
To further assess the performance of PatchRec-I in details, we compare it with TALLRec under two distinct experimental settings across three datasets, addressing the following research questions:
\begin{itemize}[leftmargin=*]
    \item \textbf{RQ1:} Does compression result in a loss of recommendation performance when processing identical interaction sequences?
    \item \textbf{RQ2:} For a fixed inference computational cost, does compression improve recommendation performance?
\end{itemize}

\subsubsection{\textbf{Performance Comparison under the Same Items in the Interaction Sequence (RQ1)}}

We maintain the same truncation length for interaction sequences, with item counts of 20, 40, 60, 80, and 100, and compare HR@20 of PatchRec-I with TALLRec to examine whether the compression of early items in PatchRec-I leads to any loss of recommendation performance.
The results are shown in Figure \ref{fig:ft1-history}.

On the MovieLens-1M and Goodreads datasets, PatchRec-I consistently outperforms TALLRec, with relative gains exceeding 5\% and 30\%, respectively. 
This improvement may be attributed to the increased information density in the compressed item patches, which mitigates the influence of noisy interactions.
On the MovieLens-100K dataset, PatchRec-I achieves performance comparable to TALLRec, indicating that the proposed compression strategy does not compromise recommendation accuracy.

Furthermore, under a fixed compression ratio, we observe that as the truncation length (\ie the items in the interaction sequence) increases across the three datasets, the recommendation performance of PatchRec-I generally improves, unlike TALLRec, which doesn't show steady improvement. 
This trend suggests that, for PatchRec-I, incorporating more historical interactions enhances the ability to model long-term user behaviors effectively.

\subsubsection{\textbf{Performance Comparison under Comparable Numbers of Tokens (RQ2)}}
We compare the recommendation performance of PatchRec-I with TALLRec with comparable token counts for LLM input to evaluate the impact of compression on recommendation performance at an equivalent computational cost.
The results are shown in Figure \ref{fig:ft1-token}.

In all experiments across the three datasets, PatchRec-I processes 100 items, which are represented as item patches or raw textual titles. 
However, due to varying compression ratios, the number of tokens required differs. 
For each bar in the figures, both PatchRec-I and TALLRec input the same number of tokens. 
Despite this parity in token usage, PatchRec-I consistently achieves higher HR@20 than TALLRec, with relative improvements of at least 7\%, 20\%, and 2\%, and reaching up to 13\%, 72\%, and 19\% on MovieLens-1M, Goodreads, and MovieLens-100K, respectively.

These results demonstrate that across all three datasets, PatchRec-I delivers superior recommendation performance compared to TALLRec at the same computational cost, underscoring the efficiency and effectiveness of the proposed compression method.

\subsection{Analysis of PatchRec-S}

PatchRec-S further compresses the earliest item patches into session patches, thereby achieving even higher compression ratio.
To validate its effectiveness, we compare the compression ratio and HR@20 of three methods --- TALLRec, PatchRec-I, and PatchRec-S --- while maintaining the same number of 100 items in the interaction sequence across different compression ratios on three datasets. 
The results are shown in Figure~\ref{fig:ft2_ml1m} and Figure~\ref{fig:ft2}.

The figures show that the distribution of PatchRec-I and PatchRec-S points is positioned further towards the upper-right corner compared to TALLRec across all three datasets. Specifically:
\begin{itemize}[leftmargin=*]
    \item PatchRec-S consistently outperforms TALLRec in HR@20 across various compression ratios on the MovieLens-1M and Goodreads datasets. 
    Even at compression ratios as high as 9.23 and 13.62 for MovieLens-1M and Goodreads, respectively, PatchRec-S maintains superior HR@20 compared to TALLRec.
    \item PatchRec-I points are generally positioned higher on the performance axis, while PatchRec-S points are shifted further to the right, reflecting greater efficiency.
\end{itemize}
These results indicate that both PatchRec-I and PatchRec-S provide a better performance-efficiency trade-off compared to TALLRec. 
Furthermore, the distinction between the two methods demonstrates their complementary strengths: PatchRec-I prioritizes performance, while PatchRec-S emphasizes efficiency.

\subsection{Effectiveness of the Patch Pre-training Stage}
In this section, we ablate to demonstrate the benefits of the patch pre-training stage.
\subsubsection{\textbf{Impact of the Patch Pre-training Stage}}
\begin{table}[t]
\centering
\caption{The performance comparison of PatchRec-I/S between with and without patch pre-training on MovieLens-1M dataset.
The patch pre-training stage improves recommendation performance.}
\vspace{-4pt}
\label{tab:ablation}
\begin{small}
\begin{tabular}{lcccc}
\toprule
\textbf{Method} & \textbf{HR@10} & \textbf{N@10} & \textbf{HR@20} & \textbf{N@20} \\
\hline\hline
\multicolumn{5}{l}{\cellcolor{gray!16}\textbf{PatchRec-I}}\\ 
\midrule
\textbf{w/o Patch Pre-training} & 0.1014 & 0.0439 & 0.1550 & 0.0504 \\
\textbf{w/ Patch Pre-training} & \textbf{0.1058} & \textbf{0.0455} & \textbf{0.1616} & \textbf{0.0525} \\
\hline\hline
\multicolumn{5}{l}{\cellcolor{gray!16}\textbf{PatchRec-S}}\\ 
\midrule
\textbf{w/o Patch Pre-training} & 0.0920 & 0.0386 & 0.1436 & 0.0463 \\
\textbf{w/ Patch Pre-training} & \textbf{0.0967} & \textbf{0.0408} & \textbf{0.1526} & \textbf{0.0496} \\
\bottomrule
\end{tabular}
\end{small}
\end{table}
To evaluate the importance of the patch pre-training stage, we compare the recommendation performance of PatchRec-I and PatchRec-S under two settings:
(1) Starting from LLM checkpoints trained with the patch pre-training stage, which is denoted as w/ patch pre-training.
(2) Directly initialized from the Llama-3.2-1B-Instruct checkpoint without patch pre-training, denoted as w/o patch pre-training.
The evaluation is conducted on the MovieLens-1M dataset, and the results are shown in Table~\ref{tab:ablation}.
From the table, we can observe that both PatchRec-I and PatchRec-S with the patch pre-training stage consistently outperforms those without that stage.

This demonstrates the critical role of the patch pre-training stage in improving performance. 
During this stage, the inclusion of both uncompressed textual tokens (\ie item titles) and compressed item patches within the same batch enables the LLM to effectively learn the patterns of item-level compression. 
This adaptation allows the LLM to transition seamlessly from the textual token space to the compressed item patch space, laying a robust foundation for the subsequent multi-grained patch fine-tuning stage.

\subsubsection{\textbf{On Data Augmentation}}
\begin{table}[t]
\centering
\caption{Performance comparison between TALLRec, patch pre-training, and data augmentation with dropout on MovieLens-1M dataset.
The enhancements achieved by compression are not simply due to the random exclusion of items.}
\vspace{-4pt}
\label{tab:dropout}
\begin{small}
\begin{tabular}{lcccc}
\toprule
\textbf{Method} & \textbf{HR@10} & \textbf{N@10} & \textbf{HR@20} & \textbf{N@20} \\
\midrule
\textbf{TALLRec} & 0.0933 & 0.0396 & 0.1482 & 0.0491 \\
\textbf{Dropout} & 0.0936 & 0.0402 & 0.1480 & 0.0492 \\
\cellcolor{gray!16}\textbf{Patch Pre-training} & \cellcolor{gray!16}\textbf{0.1026} & \cellcolor{gray!16}\textbf{0.0431} & \cellcolor{gray!16}\textbf{0.1610} & \cellcolor{gray!16}\textbf{0.0532} \\
\bottomrule
\end{tabular}
\end{small}
\end{table}
To examine the performance improvements attributed to the compression operations in the patch pre-training stage, we compare our compression approach against a baseline method that directly drops the items compressed during the patch pre-training data augmentation process.

The experimental results, as summarized in Table~\ref{tab:dropout}, reveal that patch pre-training consistently outperforms both the TALLRec (\ie no-compression) and dropout baselines. 
This demonstrates that the enhancements achieved by our compression strategy are not simply due to the random exclusion of items.
Our compression design retains critical patterns while increasing the effective information density of the input.
These findings highlight the strength of our method in balancing efficiency and effectiveness, ensuring that compression enhances.

\subsection{Performance and Efficiency for Long-term User Behaviors}
\begin{figure}[t]
\centering
\includegraphics[width=0.48\textwidth]{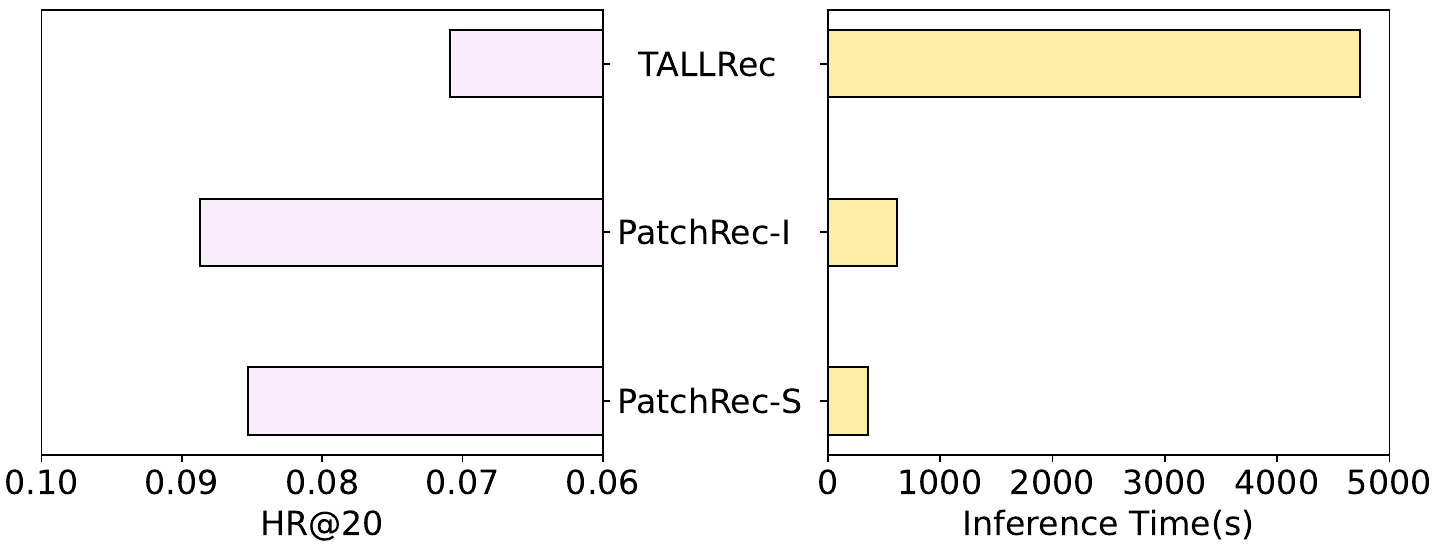}
\vspace{-8pt}
\caption{Performance (\ie HR@20) and efficiency (\ie inference time) comparison of TALLRec, PatchRec-I, and PatchRec-S for long sequences (\ie users with at least 500 behaviors) on MovieLens-1M.
PatchRec enhances both performance and efficiency in modeling long-term user behaviors.}
\vspace{-4pt}
\label{fig:long}
\end{figure}

To further evaluate the performance and efficiency of PatchRec when applied to even longer sequences, we analyze user groups with more than 500 interactions in MovieLens-1M.
This experiment was conducted using a Nvidia A100 GPU with 40 GB of memory.
For each test interaction, we maximize the number of historical items visible to all three methods (\ie TALLRec, PatchRec-I, and PatchRec-S) to fully utilize the available GPU memory, which simulates the pragmatism in real-world recommender systems that make full use of all available computing sources and data.
Specifically, we truncate the Latest-$K$ ($K$=150, 350) historical items for both TALLRec and PatchRec-I, whereas PatchRec-S is capable of accommodating all historical items (1435 at most) within the 40 GB GPU memory constraints.

In the case of PatchRec-I, the latest 20 interactions are retained as textual titles, while earlier interactions are compressed into item patches.
For PatchRec-S, the parameter $L$, which denotes the number of items in a session, is set to 20.
The experimental results are shown in Figure \ref{fig:long}.
The figure demonstrates that PatchRec-I achieved a 25.11\% improvement in HR@20 with only 12.99\% inference time, compared to TALLRec. 
Similarly, PatchRec-S achieved a 20.31\% increase in HR@20 relative to TALLRec, with a even higher 13.09-fold reduction in inference time. 
These results indicate that our proposed methods significantly enhance both performance and efficiency in modeling long-term user behaviors, implying that our PatchRec is capable of modeling practical historical behavior lengths via LLM4Rec.

\section{Related Work}
\subsection{LLM for Recommendation}
With LLMs demonstrating vast world knowledge and impressive reasoning capabilities across various domains \cite{gpt-4, gemini, llama-3, qwen2, deepseek, patch-pretrain}, their potential for recommendation tasks has garnered significant attention from researchers \cite{zero-shot-ranker, llamarec}.
Existing work on LLM4Rec can be roughly classified into two categories:
(1) LLMs as enhancers: LLMs augment traditional recommender models by processing textual features associated with users and items \cite{morec, unisrec}, and (2) LLMs as recommenders: LLMs directly generate recommendations for users' next interactions \cite{chatrec, tallrec, tiger, p5}.

The primary focus of this paper is on the latter --- LLMs as recommenders --- where the core challenge is converting user-item interaction sequences into a format suitable for language modeling.
Current item indexing approaches in LLM4Rec primarily fall into two types:
(1) Text-based representations, where items are indexed using textual information, such as item titles \cite{tallrec, bigrec}, descriptions \cite{recformer}, or learned semantic IDs \cite{lmindexer, p5_index};
(2) Collaborative-incorporated representations, which consider collaborative signals \cite{collm, llara} in item embeddings.

For both aforementioned types, each item is typically represented by multiple tokens. 
Due to the quadratic computational complexity $O(n^2)$ of self-attention mechanism in LLMs \wrt sequence length~\footnote{The computational complexity of vanilla attention mechanisms is $O(n^2)$. 
Recent studies proposed methods to reduce it to approximately $O(n)$}, using multiple tokens to represent one item in a user's interaction sequence becomes computationally demanding. 
This motivates us to explore more efficient item and even sequence representations, aiming to improve the performance-efficiency trade-off. 

\subsection{Long Sequence Modeling in LLM4Rec}
Recent efforts have made significant strides in addressing the context length limitations of LLMs \cite{ct,lost,bigbird,rope,longrope}. 
One prominent line of work focuses on context compression, wherein older or less critical information is compressed into a more compact representation.
Early method \cite{AC} trains LLMs to summarize or encode lengthy preceding text into compact vectors that can be prefixed to the prompt.
Similarly, ICAE \cite{ICAE} introduces an in-context autoencoder that encodes long contexts and then reconstructs necessary details when needed.
More recent studies \cite{gist,compact} explore prompt compression, demonstrating that selectively retaining only task-relevant facts can preserve performance while substantially reducing input length.
Broadly, these compression-based techniques aim to prevent earlier information from overwhelming the model’s capacity as context grows.
In contrast to these approaches, PatchRec learns concise embeddings at multiple granularities, enabling efficient modeling of extended histories without relying on explicit textual compression or summarization.

The integration of LLMs into recommendation systems introduces further challenges, particularly the restriction on modeling long user behavior sequences due to limited context length and computational resource constraints \cite{llmrank,r3mem,sr-survey,llm-cr}.
One solution \cite{pure,memorybank} enhances LLMs with external long-term memory modules that summarize user interaction histories and maintain dynamically updated profiles.
Rather than inputting every past interaction, MemoryBank \cite{memorybank} maintains a running summary that the model can consult when needed.
PURE \cite{pure} introduces a ``Review Extractor'' and ``Profile Updater'' continuously distill evolving user histories into a compact form.
Another notable approach, ReLLA \cite{rella} tackles the challenge by retrieving the top-$K$ interacted items from a user's history based on semantic relevance to the target item. 
Distinct from these memory- and retrieval-based methods, PatchRec directly trains LLMs to internalize long-term historical signals through aggregated embeddings, eliminating the need for explicit summarization or retrieval mechanisms.
\section{Limitation}
While PatchRec shows a strong performance-efficiency trade-off, several limitations remain, especially for real-world deployment:
\begin{itemize}[leftmargin=*]
    \item \textbf{Inference Latency.} Despite compression, reliance on LLMs still incurs high inference costs, limiting real-time applicability. Future work can explore model distillation, quantization, or early-exit strategies to reduce latency.
    \item \textbf{Compression Design.} PatchRec is simple and effective, but more expressive designs --- such as advanced patching or innovative grouping techniques --- could be further investigated.
    \item \textbf{Adaptive Granularity.} We currently use a fixed multi-grained compression scheme. Dynamically adjusting granularity based on resource constraints or user patterns is a promising direction.
\end{itemize}
Addressing these limitations will be key to making PatchRec more practical and scalable for real-world recommendation systems.
\section{Conclusion}
In this paper, we introduce PatchRec, a simple yet effective compression framework designed for multi-grained modeling of users' extensive historical behaviors in LLM-based recommendation (LLM4Rec).
To address the significant challenges posed by LLM constraints and the temporal dynamics of user interactions --- mostly untouched in prior works --- PatchRec leverages a hierarchical and time-aware compression strategy, enabling a deeper and more nuanced understanding of user preferences.
By adopting a two-stage training process, PatchRec effectively adapts LLMs to operate within a multi-grained sequence representation space.
Extensive experiments on benchmark datasets demonstrate that PatchRec not only dramatically reduces the input token length for LLMs but also consistently improves recommendation performance.

This work paves the way for future research into more efficient and effective ways to harness the power of LLMs in sequential recommendation.
We hope PatchRec will inspire further exploration into efficient LLM-based sequential recommenders for lifelong user sequence modeling.

\begin{acks}
This research is supported by the National Natural Science Foundation of China (92270114, U24B20180, 62121002) and the Young Elite Scientists Sponsorship Program by CAST (2023QNRC001).
\end{acks}

\bibliographystyle{ACM-Reference-Format}
\balance
\bibliography{sample-base}

\end{document}